\documentclass[12pt,english]{article}
  \usepackage[T1]{fontenc}
  \usepackage{lmodern}
  \usepackage[utf8]{inputenc}
  \usepackage{geometry}
  \usepackage{amsthm}
  \usepackage{amsmath}
  \usepackage{amssymb}
  \usepackage{graphicx}
  \usepackage{setspace}
  \usepackage{color}
  \usepackage{braket}
  \usepackage{enumerate}	
  \usepackage{microtype}
  \usepackage{fancyhdr}
  \usepackage{titling}
  \usepackage[round]{natbib}
  
\usepackage{footmisc}

\usepackage[round]{natbib}

\usepackage{sectsty}
\subsectionfont{\normalfont\itshape} 
\begin{document}
\pagestyle{empty}
\title{On the Equivalence of von Neumann and Thermodynamic Entropy}
\author{Carina E. A. Prunkl\\ \textit{Future of Humanity Institute, University of Oxford}}
\date{}% left justifying the text
\maketitle
\raggedright
\onehalfspacing

\section*{Abstract}

In 1932, von Neumann argued for the equivalence of the thermodynamic entropy and $-\text{Tr}\rho\ln\rho$, since known as the von Neumann entropy. Hemmo and Shenker (2006) recently challenged this argument by pointing out an alleged discrepancy between the two entropies in the single particle case, concluding that they must be distinct. In this article, their argument is shown to be problematic as it a) allows for a violation of the second law of thermodynamics and b) is based on an incorrect calculation of the von Neumann entropy. 

%I will restore the analogy by identifying a series of common misconceptions in Hemmo and Shenker's argument, which are typically made in single particle scenarios and regard the classification of entropy in the presence of an external, correlated agent, leading to all sorts of peculiar consequences.

\newpage
\pagestyle{plain}
\singlespacing
\tableofcontents
\newpage 

\section{Introduction}

In \textit{Mathematische Grundlagen der Quantenmechanik} von Neumann introduces $-\text{Tr}\rho\ln\rho$ as the quantum mechanical generalisation of the phenomenological thermodynamic entropy, where $\rho$ is the quantum mechanical density operator\footnote{Von Neumann considers two types of processes that describe how the quantum state changes in time. The first, `Prozess 1', is associated with the probabilistic outcome of a measurement, whereas `Prozess 2' refers to the evolution of the system via the Schr\"odinger equation. $-\text{Tr}\rho\ln\rho$ is shown to be non-decreasing for both of them.}. In his argument, he considers the cyclic transformation of a quantum gas confined to a box. By demanding that the overall entropy change of system and heat bath must be zero by the end of the cycle\footnote{An explicit assumption of the validity of the second law.}, von Neumann concludes that the entropy of the quantum gas ought to be given by $-\text{Tr}\rho\ln\rho$. Hemmo and Shenker (2006) recently challenged this argument by pointing out an alleged discrepancy between the two entropies in the single particle case, concluding that they must be distinct. In this article I demonstrate that their argument against the equivalence of thermodynamic and von Neumann entropy is problematic as it a) allows for a violation of the second law of thermodynamics and b) is based on an incorrect calculation of the von Neumann entropy. The article is structured as follows: after a summary of von Neumann's original argument, I will quickly revisit the debate that has been lead to date by \cite{shenker_is_1999} and \cite{henderson_von_2003} before moving on to an analysis of Hemmo and Shenker's (2006) most recent contribution, which will be shown to be problematic.
%Structurally similar to the classical Gibbs entropy \citep{gibbs_equilibrium_1878}, 
%
%In his seminal book, \cite{von_neumann_mathematische_1996} presents an argument in which he determines the entropy of a quantum mechanical ensemble with density matrix $\rho$ and establishes that entropy is non-decreasing

\section{The Argument}

To fully appreciate Hemmo and Shenker's criticism, it is helpful to begin by recapitulating von Neumann's argument as presented in his \textit{Mathematische Grundlagen}. 
\medskip

\subsection{The Setup}

Von Neumann begins by considering $N$ non-interacting, quantum systems, denoted by $\mathbf{S}_1,...,\mathbf{S}_N$. For the purpose of this paper, we take these systems to be two-state quantum systems and only consider, say, the spin states of a spin-$\frac{1}{2}$ particle.\footnote{A generalisation to more degrees of freedom is straight forward and may be found in \citep[pp.191--201]{von_neumann_mathematische_1996}.} Each of these systems is placed in a box $\mathbf{K}_i$, with $i=1,...N$, whose walls shield its contained system off from its environment and thus prevent any interaction between systems. The boxes $\mathbf{K}_1,...,\mathbf{K}_N$ are now all placed into another, much larger box $\bar{\mathbf{K}}$ of volume $V$. For simplification, von Neumann assumes that there are no force fields, in particular no gravitational fields, present in $\bar{\mathbf{K}}$. This in turn means that there is no gravitational interaction between the boxes $\mathbf{K}_1,...,\mathbf{K}_N$, even though they may exchange kinetic energy via collisions. The boxes can thus be taken to behave just like molecules of a gas. Von Neumann calls this `quantum gas' a  $[\mathbf{S}_1,...,\mathbf{S}_N]$-gas, where $[\mathbf{S}_1,...,\mathbf{S}_N]$ is the statistical ensemble associated with the systems. A more detailed analysis of von Neumann's understanding of such an ensemble, or \textit{Gesamtheit}, will be presented in Section \ref{sec:response}. The large box $\bar{\mathbf{K}}$ can be brought into thermal contact with a heat bath at temperature $T$ and in that case, after some equilibration time, the $[\mathbf{S}_1,...,\mathbf{S}_N]$-gas itself will be at temperature $T$.\footnote{For a more detailed account of this equilibration process, see \citep[pp.192--193]{von_neumann_mathematische_1996}.} Finally, we add an \textit{empty} box $\bar{\mathbf{K}}^\prime$ of equal volume $V$ to the right of $\bar{\mathbf{K}}$. 
\medskip

\begin{figure}[h!]
\hspace*{3cm}
\includegraphics{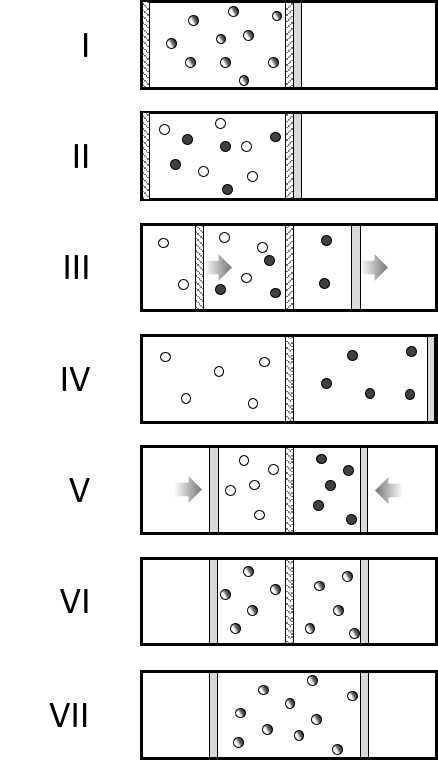}
\caption{\small An illustration of von Neumann's argument. \textbf{(I)} The individual two-state quantum systems are each in state $\ket{0}=w_1\ket{+}+w_2\ket{-}$, indicated by bicoloured circles. For illustrative purposes, we assume $w_1=w_2=\frac{1}{\sqrt{2}}$. \textbf{(II)} After the measurement in the $\ket{+}/\ket{-}$-basis, the system is now described by a statistical mixture of $\ket{+}$- (white) and $\ket{-}$-systems (black). \textbf{(III)} The $\ket{+}$- and $\ket{-}$-systems are now separated by using semi-permeable membranes to 'push' the $\ket{-}$-systems into the right box $\bar{\mathbf{K}}$. \textbf{(IV)} The $\ket{+}$- and $\ket{-}$-systems are now completely separated. \textbf{(V)} The two boxes are compressed to half of their respective volumes. \textbf{(VI)} The $\ket{+}$- and $\ket{-}$-gases are now transformed back into their original state of superposition. \textbf{(VII)} The partition is removed. In this figure, no heat bath is present, but we assume that it exists in the background and takes up the dissipated entropy at $4\rightarrow 5$.}
\label{shenkerfig1}
\end{figure}

\subsection{The Process}
 
The $[\mathbf{S}_1,...,\mathbf{S}_N]$-gas now undergoes a series of transitions. The following various stages are also illustrated in Figure \ref{shenkerfig1}.\footnote{The presentation of the argument given here slightly differs from its original. Instead of having the gas undergo a cyclic process (in Figure \ref{shenkerfig1}, the system is in the same state at Stages I and VII), von Neumann separately considers the system's entropy changes for Stages II to VII \citep[pp.200--202]{von_neumann_mathematische_1996} and Stages I to II \citep[pp.202--206]{von_neumann_mathematische_1996}. The cyclic version was chosen to be consistent with the presentation of von Neumann's argument in \citep{shenker_is_1999,henderson_von_2003} and \citep{hemmo_von_2006}.}

\textbf{Stage I}: Each of the two-level quantum systems $\mathbf{S}_i$ initially is in the pure state $\ket{0}=w_1\ket{+}+w_2\ket{-}$, where the states $\ket{+}$ and $\ket{-}$ can be taken to be spin eigenstates of a spin-half particle. Given the lack of interaction between the quantum systems, the density matrix of the overall system factorises as $\rho=\rho_1\otimes...\otimes\rho_N$ with $\rho_i=\ket{0}\!\bra{0}$. 

\textbf{Stage II}: Each system is now measured in the \{$\ket{+}\!,\!\ket{-}$\}-basis, resulting in a mixture with $w_1^2N$ particles in state $\ket{+}$ and $w_2^2N$ particles in state $\ket{-}$.

\textbf{Stage III \& IV}: The $\ket{+}$- and $\ket{-}$-systems are now separated in the following manner. The wall between $\bar{\mathbf{K}}$ and $\bar{\mathbf{K}}^\prime$ is replaced with a \textit{movable} partition and a \textit{fixed} semipermeable membrane\footnote{For such a semipermeable membrane to exist in principle, the two states need to be orthogonal. Mixtures instead of pure states are also conceivable, as long as they are disjoint.}. This first semipermeable membrane is transparent to $\ket{-}$-systems but impermeable to $\ket{+}$-systems. From the very left of $\bar{\mathbf{K}}$, another semipermeable membrane is inserted. It is movable and furthermore transparent to $\ket{+}$-systems, while being impermeable to $\ket{-}$-systems. The $\ket{-}$-systems are now 'pushed' into the right box $\bar{\mathbf{K}}^\prime$ by moving this second semipermeable membrane and the movable partition in the center \textit{simultaneously} and quasi-statically to the right, keeping the enclosed volume constant at all times (see Figure \ref{shenkerfig1}, Stage III, for an illustration of this step). During this process, no work is done on the gas and no heat is exchanged with the heat bath. Eventually, all the $\ket{-}$-systems will be in $\bar{\mathbf{K}}^\prime$, while the $\ket{+}$-systems remain in the left box (Figure \ref{shenkerfig1}, Stage IV).

\textbf{Stage V}: The two boxes are now isothermally compressed to volumes $w_1^2V$ and $w_2^2 V$. Figure \ref{shenkerfig1} illustrates this step for $w_1^2=w_2^2=\frac{1}{2}$. The particle densities in $\bar{\mathbf{K}}$ and $\bar{\mathbf{K}}^\prime$ change from $w_1^2N/V$ and $w_2^2N/V$ to $N/V$ where, as before, $N$ is the total number of systems. The entropy of the heat reservoir increases by $Nk_B w_1^2\ln w_1^2$ and $Nk_B w_2^2\ln w_2^2$ respectively, where $k_B$ is the Boltzmann constant.

\textbf{Stage VI}: The $\ket{+}$- and $\ket{-}$-gases are then reversibly transformed back into a $\ket{0}$-gas via unitary operations. 

\textbf{Stage VII}: Finally, the partition between the two chambers is removed, restoring the original state of the gas. 

Von Neumann's now argues as follows: all transitions between Stages II and VII take place in a reversible fashion, which means that the total entropy change of gas and heat bath between II and VII must be zero. Since a total amount of $\Delta S=Nk_B\left[ w_1^2\ln w_1^2+w_2^2\ln w_2^2\right]$ has been dumped into the heat bath during the compression stage, and since the (normed) entropy of the final $\ket{0}$-gas is zero by definition, the entropy of the gas must have been $S=S_{+}+S_{-}=-Nk_B\left[ w_1^2\ln w_1^2+w_2^2\ln w_2^2\right]$ before. Later in his book, von Neumann explains that the measurement process ('Prozess 1') is responsible for the entropy increase between Stages I and II \citep[pp.202--206]{von_neumann_mathematische_1996}.

The above considerations are easily generalised to more dimensions: for a system described by a density matrix $\rho$ with eivenvectors $\ket{\phi_1},...,\ket{\phi_n}$ and eigenvalues $w_1,...,w_n$, the entropy is then given by $S_\rho=-\text{Tr}\rho\ln\rho=-\sum_{i=1}^n w_i\ln w_i$.

%Von Neumann's argument is designed to show that, by exclusion, the irreversible step of what he calls a Type measurement is associated with an increase in entropy.

\section{\citeauthor{shenker_is_1999}'s Criticism and \citeauthor{henderson_von_2003}'s Reply
}

I will now briefly consider Shenker's (1999) first criticism against von Neumann's argument and Henderson's (2003) reply. According to Shenker, two assumptions were made by von Neumann: 

\begin{enumerate}[a)]
	
	\item  the thermodynamic entropy only changes during the compression, stages IV to V, and 
	\item the entropies of stage I and VII are the same.
\end{enumerate} 

As Shenker presents the argument, von Neumann's conclusion was that the entropy must thus have increased during the measurement process (I to II), to balance out the decrease during the compression (IV to V). 

In order to show that von Neumann entropy and ``classical entropy''\footnote{Shenker does not distinguish between classical statistical mechanical entropy and thermodynamic entropy at this point: ``Classical thermodynamics concludes that the very separation [of two gases] means a reduction of entropy. This is called entropy of mixing [...]'' \citep[39]{shenker_is_1999}. Entropy of mixing however has its origin in statistical and not phenomenological considerations.} are distinct, Shenker points out an alleged discrepancy in behaviour, which supposedly takes place between stages II and IV. Following her argument, let us first consider the change in von Neumann entropy: at stage II, the system is in a mixed state and has positive von Neumann entropy. At stage IV, the system is in a pure state and, so Shenker claims, has zero von Neumann entropy by definition. The von Neumann entropy therefore must have decreased between II and IV, that is, it must have decreased during the separation of the $\ket{+}$- and $\ket{-}$-systems.

``From a thermodynamic point of view'' (p.42), however, the entropy has not changed between II and IV. This is because ``the entropy reduction of the separation is exactly compensated by an entropy increase due to expansion'' \citep[p.45]{shenker_is_1999}. The thermodynamic entropy instead changes between during the compression, IV to V. According to Shenker,  
Thermodynamic entropy and von Neumann entropy therefore differ in their behaviour since the reduction in thermodynamic entropy takes place at a later stage (IV to V) than the reduction of von Neumann entropy (II to IV).

\cite{henderson_von_2003} points out some deficiencies in Shenker's argument that explain the alleged discrepancy. She shows that the system at stage IV \textit{cannot} be considered to be in a pure state, as the gas' spatial degrees of freedom ought also be taken into account in addition to its spin degrees of freedom. The initial state of the system at stage I is then given by $\ket{0}\otimes \rho_{\beta}$, where $\rho_{\beta}$ is the thermal state of the system in contact with a heat bath at inverse temperature $\beta$. The entropy change at stage II is then only due to the entropy change of the spin degrees of freedom. Furthermore, even if we assume collapse, as Shenker implicitly does, the entropy is still high, since ``we lack knowledge of \textit{which} pure state the system is in'' \citep[p.294, original emphasis]{henderson_von_2003}. The separation step between II and IV then only `labels' the states in so far as they are associated with a particular spatial area of the box, but this step does \textit{not} change the entropy. The change in entropy at the compression stage V is then due to a change of the entropy of the \textit{spatial} degrees of freedom.

\section{Modern Criticism by \citeauthor{hemmo_von_2006}}

In a subsequently published, revised and amended version, \cite{hemmo_von_2006} offer an amended proposal with a similar but slightly weakened claim. They assert that ``von Neumann's argument does not establish a conceptual link between $-\text{Tr}[\rho\ln\rho]$ and the thermodynamic quantity $(1/T)\int pdV$ (or $dQ/T$) \textit{in the single particle gas} [...]'' \citep[p.158, emphasis added]{hemmo_von_2006}. They therefore retain their position that the von Neumann entropy cannot be empirically equivalent to the phenomenological entropy, but restrict this inequivalence to the domain of single or sufficiently few particles. Von Neumann and thermodynamic entropy, they argue, are effectively equivalent only in the thermodynamic limit.

This section will discuss Shenker's and Hemmo and Shenker's (H\&S) efforts to show dissimilar behaviour between the two entropies and reveal that their argument is problematic to the extent that it allows for \textit{pepetua mobile} of the second kind. The shortcoming will be identified as the failure to take into account the entropy contribution of the measurement apparatus. I will make use of the famous one-particle engine developed by Szilard's (\citeyear{szilard_decrease_1929}) in order to show that measurement based correlations with an external agent cannot be ignored in the single particle limit, as they straightforwardly lead to a violation of the second law. Once the entropy contribution of the measurement apparatus is taken into account, however, the analogous behaviour of thermodynamic entropy and von Neumann entropy for the \textit{joint system} is restored.

%   \cite{hemmo_von_2006} ignore these contributions and therefore arrive at erroneous results\footnote{It should be pointed out that even by showing that Hemmo's and Shenker's argument does not hold, this does not imply the claim that von Neumann's argument established thermodynamic entropy and von Neumann entropy indeed \textit{are} equivalent, after all they have distinct domains of application. However, to be a quantum statistical mechanical counterpart to the phenomenological entropy, the von Neumann entropy better behaves like the thermodynamical entropy in all relevant, i.e. measurable instances.}. 

\begin{figure}[h!]
\centering
\includegraphics[width=.5\textwidth]{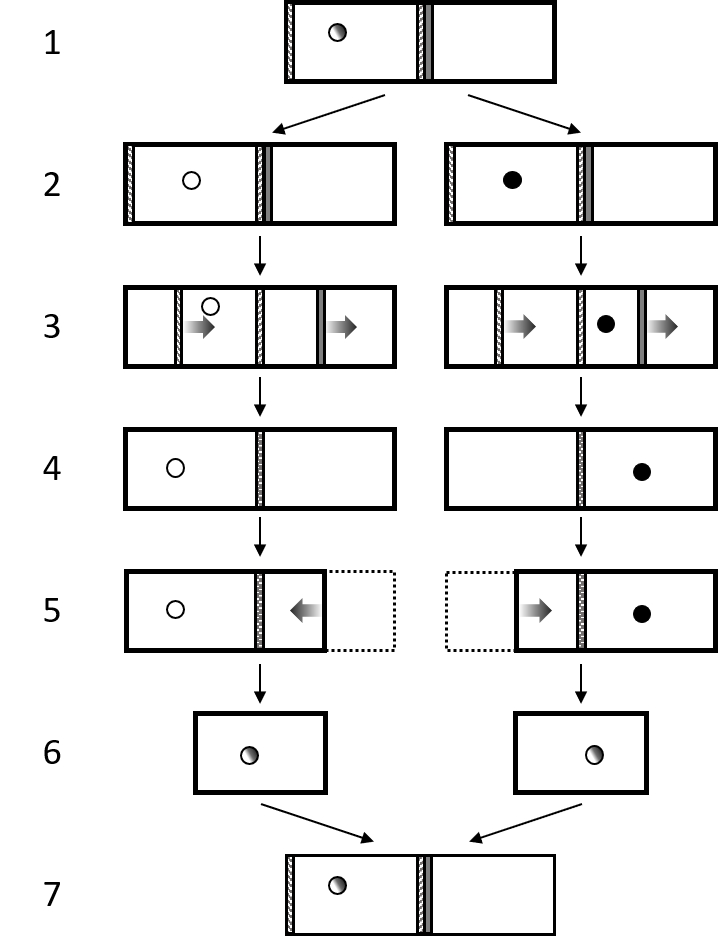}
\caption{\small Illustration of the Gedankenexperiment following \cite{hemmo_von_2006}. (1) the particle is prepared in a spin-$x$ up eigenstate. (2) A spin-$z$ measurement is performed on the particle. (3) Depending on the outcome, the particle is moved to the right side of the box or remains in the left side via semi-permeable membranes. (4) A location measurement is performed. (5) The empty side of the box is compressed. (6) \& (7) The system is brought back to its original state.}
\label{fig:singleparticle}
\end{figure}

The following argument, including any conceptual ambiguities, has been taken unamended from \citep{hemmo_von_2006}, with the exception that we represent the position of the particle by the two orthogonal states $\ket{L}$ and $\ket{R}$, where $L$ and $R$ stand for 'Left' and 'Right', as opposed to Hemmo and Shenker's mixed state $\rho(L)$ representation. For the remainder of the article I furthermore assume, just like H\&S, that it is in fact possible to treat a single quantum particle as a genuine thermodynamic system. An illustration of the following steps can be found in Figure \ref{fig:singleparticle}.

\textbf{Step 1 (Preparation I)}: A quantum particle $P$ is prepared in a spin-up eigenstate in the $x$-direction, $\ket{+_x}_P$. Its initial location is given by $\ket{L}_P$, where $L$ and $R$ refer to its position in either the left or the right part of the box. The measuring apparatus $M$ starts out in the state $\ket{\text{Ready}}_M$. The initial state of the particle is then given by the product state:

\begin{equation}
\rho^{(1)}=\ket{+_x}\!\bra{+_x}_P\ket{L}\!\bra{L}_P\ket{\text{Ready}}\!\bra{\text{Ready}}_M.
\end{equation}

\textbf{Step 2 (Preparation II)}: In Step 2, a measurement in the spin $z$ direction is performed, leading to an entanglement of the measurement apparatus' pointer states and the $z$ spin eigenstates. It is important to note that H\&S do not specify the nature of the measurement at this stage, i.e. whether they are working in a collapse or no-collapse model. The state of the entire system is given by 

\begin{equation}
\rho^{(2)}=\frac{1}{2}\left( \ket{+_z}\!\bra{+_z}_P\ket{+}\!\bra{+}_M+\ket{-_z}\!\bra{-_z}_P\ket{-}\!\bra{-}_M\right)\ket{L}\!\bra{L}_P,
\end{equation}

while the reduced density matrix of the particle becomes:

\begin{equation}
\rho^{(2,red)}=\frac{1}{2}\left( \ket{+_z}\!\bra{+_z}_P+\ket{-_z}\!\bra{-_z}_P\right)\ket{L}\!\bra{L}_P, \label{eq:step2}
\end{equation}

which, as H\&S state, ``in some interpretations may be taken to describe our ignorance of the z spin of P'' \citep[p.160]{hemmo_von_2006}. 

Whereas the von Neumann entropy of the spin component $S_{vN}=-\text{Tr}[\rho\ln\rho]$ was zero before, it now becomes positive. The thermodynamic entropy however, the authors assert, remains the same.

\textbf{Step 3 (Separation)}: Two semi-permeable membranes are inserted and moved through the box in such a way that the particle remains on the left if it is in state $\ket{+_z}$ but is moved to the right if it is in state $\ket{-_z}$. There is no work cost involved in this process and neither von Neumann entropy nor thermodynamic entropy change during this step, during which the spatial degrees of freedom are coupled to the spin degrees of freedom.

%The state of the system after the separation is:
%\begin{equation}
%\rho^{(3,red)}=\frac{1}{2} %\ket{+_z}\bra{+_z}_P\ket{L}\!\bra{L}_P %+\frac{1}{2}\ket{-_z}\bra{-_z}_P\ket{R}\!\bra{R}_P.
%\end{equation}

\textbf{Step 4 (Measurement)}: 
As we are only considering a \textit{single} molecule in this setup as opposed to von Neumann's original many particle gas, the compression stage needs to be preceded by a location measurement in order to determine which part of the box is empty. H\&S therefore introduce a \textit{further} measurement before compression (not present in von Neumann's original argument), in order to determine in which part of the box the particle is located.

H\&S add that for the calculation of the von Neumann entropy, collapse and no-collapse interpretations will now have to use different expressions for the quantum state. In collapse theories, the state as a result of the location measurement collapses into either

\begin{equation}
\rho^{(4,+)}=\ket{+_z}\bra{+_z}_P\ket{L}\!\bra{L}_P\;\;\;\text{or}\;\;\;\rho^{(4,-)}=\ket{-_z}\bra{-_z}_P\ket{R}\!\bra{R}_P,.\label{collapse}
\end{equation}

For no-collapse interpretations on the other hand, the system's state is given by the reduced density matrix:

\begin{equation}
\rho^{(4,red)}=\frac{1}{2} \ket{+_z}\bra{+_z}_P\ket{L}\!\bra{L}_P +\frac{1}{2}\ket{-_z}\bra{-_z}_P\ket{R}\!\bra{R}_P.\label{nocollapse}
\end{equation}

The thermodynamic entropy, $S_{TD}$, as the authors stress, is \textit{not} influenced by the position measurement and does not change during this step, in the sense--presumably--that no heat flows into, or out, of the system in consequence of this measurement . By contrast they urge, whether the von Neumann entropy changes, depends on whether we consider collapse or no-collapse interpretations. In the case of collapse interpretations the von Neumann entropy of the system allegedly decreases, whereas in the case of no-collapse interpretations, it remains the same.

\textbf{Step 5 (Compression)}: The box is isothermally compressed back to its original volume $V$. 
The change in \emph{thermodynamic} entropy during this step is normally given by $S_{TD}=(1/T)\int{pdV}$, however, since there is no work involved in the compression against the vacuum, H\&S argue, the thermodynamic entropy does not change at Step 5. In fact, the thermodynamic entropy does not change throughout the \textit{whole experiment}, the authors claim. 

\textbf{Step 6 (Return to Initial State)}: The system is brought back to its initial state by unitary transformations with no entropy cost. ``[...] [T]he measuring device need also be returned to its initial ready state. One can do that unitarily.'' \citep[161]{hemmo_von_2006}.

H\&S's main criticism thereby focuses on the fact that the thermodynamic entropy remains constant \textit{throughout the experiment}, whereas the von Neumann entropy does not:

\begin{quotation}
\small
\noindent
Therefore, whatever changes occur in $\text{Tr}\rho\ln\rho$ during the experiment, they cannot be taken to compensate for $(1/T)\int pdV$ since the latter is null throughout the experiment. \citep[p.162]{hemmo_von_2006}
\end{quotation}

\section{Discussion of \citeauthor{hemmo_von_2006}'s (\citeyear{hemmo_von_2006}) Argument}

This section will discuss the argument presented above and identify two problems. The first concerns a wrong calculation of the von Neumann entropy during the Step 4 location measurement. The second problem regards the suggested unitary reset of the measurement apparatus. %If such a unitary reset were indeed possible, then the (orthodox and probabilistic) second law could be violated on a regular basis. 

\subsection{Redundancy of the Step 4 Location Measurement}
I will begin the discussion of the above with some general observations, in order to provide some more clarity. For this, we recall that according to Hemmo and Shenker, the only difference between their and von Neumann's original thought experiment is that a \textit{further} measurement, a location measurement (Step 4), is needed to determine the molecule's location prior to the compression stage. For gases at the thermodynamic limit, this measurement becomes redundant, since the amount of molecules on each side of the box becomes proportional to their respective occupying volume. Not so for single molecules, for which, before the empty side of the box can be compressed (with probability one), a location measurement is required in order to determine which side the particle is on.

Contrary to H\&S's assertions, however, the location measurement during Step 4 is \textit{not} needed. A spin $z$ measurement already took place at Step 2 and the outcome of this measurement will be fully correlated with the position of the particle after the separation in Step 3. And so instead of introducing yet another auxiliary system that performs a location measurement on the particle, it would have been sufficient to read out the measurement result of the spin $z$ measurement.

%In fact, we have \textit{already} performed a spin measurement in Step 2, when measuring the spin $z$ direction of the system as a preparation procedure. In collapse interpretations, the system at this stage can be found in either $\ket{+_z}_P$ or $\ket{-_z}_P$. It will have already collapsed at this stage and so all further correlations are purely classical.

In the case of collapse, for example, the particle will have already collapsed into a spin eigenstate during the Step 2 measurement. The correlations established during the location measurement will thereby all be classical and reading out the spin-measurement result is sufficient to predict the particle's location after the separation. In the case of no-collapse interpretations, system and (spin-)measurement apparatus become entangled during Step 2:

\begin{equation}
\ket{\Psi}^{(2)}=\frac{1}{\sqrt{2}}\left( \ket{+_z}_P\ket{+}_M+\ket{-_z}_P\ket{-}_M\right)\ket{L}_P
\end{equation}

During the separation in Step 3, the particle's spatial degree of freedom becomes entangled with its spin degree of freedom. This means that the state of the overall system is:

\begin{equation}
\ket{\Psi}^{(3)}=\frac{1}{\sqrt{2}}\left( \ket{+_z}_P\ket{+}_M\ket{L}_P+\ket{-_z}_P\ket{-}_M\ket{R}_P\right).
\end{equation}

Therefore, for both collapse and no-collapse cases it is in fact sufficient to read out the measurement result of the Step 2 spin measurement in order to determine the location of the particle after the separation process. 

Having two instead of one measurement(s) would not be much of a problem, if it weren't the case that for H\&S, the two measurements have different consequences for the von Neumann entropy. This is the first inconsistency in their argument: whereas H\&S agree that after the spin z measurement at Step 2 the \textit{post spin measurement} density matrix of the particle is given by 

\begin{equation}
\rho^{(2,red)}=\frac{1}{2}\left( \ket{+_z}\!\bra{+_z}_P+\ket{-_z}\!\bra{-_z}_P\right)\ket{L}\!\bra{L}_P,
\end{equation}

they do not apply the same reasoning to the \textit{post location measurement} state of the system at Step 4. Instead, they use a `collapsed' density matrix to calculate the von Neumann entropy:

\begin{equation}
\rho^{(4,+)}=\ket{+_z}\!\bra{+_z}_P\ket{L}\!\bra{L}_P\;\;\;\text{or}\;\;\rho^{(4,-)}=\ket{-_z}\!\bra{-_z}_P\ket{R}\!\bra{R}_P.
\end{equation}

In the first case, the (spin) measurement has therefore increased the entropy, whereas in the second case the (location) measurement has effectively reduced it. What is going on? 

Let me first try to assemble what the authors themselves could have had in mind. In von Neumann's original argument, the Step 2 spin measurement is \textit{non-selective}\footnote{Or rather should have been, given that von Neumann himself begins his argument with a spin mixture. This however does not matter for conceptual purposes.}, which means that even if the system has \textit{de facto} collapsed into one of its eigenstates, an external agent\footnote{Some words of clarification regarding my use of the term `agent': an agent does or course not need to be a human being but can be anything that is able to measure and react to the measurement outcome accordingly. For this reason I will use the term `agent' interchangeably with the term `measurement apparatus' or even `memory cell', implying that even a simple binary system can serve as an `agent'. Using the term `agent' in this way therefore does not imply subjectivity of any kind.} would not be able to determine into which state the system has collapsed and would therefore describe the system by a density matrix $\rho^{(2,red)}$. The system would be in a so-called proper mixture, meaning that it is possible to understand $\rho$ as representing a probability distribution over pure states\footnote{In the case of no-collapse interpretations and ignoring decoherence, the agent is not yet entangled with the system. She only becomes entangled once she performs the Step 4 location measurement.}. The von Neumann entropy of the system at Step 2 has thereby increased compared to its previous state, in agreement with what von Neumann considers being the irreversibility of a  `Prozess 1'. 

The Step 4 location measurement on the other hand is \textit{selective} --- it establishes correlations between the agent who performs the measurement\footnote{More precisely the location measurement apparatus, but I will take those two to be synonymous for the time being.} and the system. These correlations then allow the agent to perform further operations on the system, such as the Step 5 compression of the box. For H\&S, the von Neumann entropy of the system at Step 4 has therefore decreased from Step 3.
%To summarise the above: In the Step 2 spin $z$ measurement the agent is not \textit{part} of the experiment, i.e. not correlated with the measurement outcomes. The state of the system is hence given by a mixed state. The agent experiences an increase in entropy. In the Step 4 location measurement on the other hand, the agent herself performs a measurement and is now \textit{correlated} with the outcome. She thereby experiences a \textit{decrease} in entropy relative to herself\todo{is it smart to mention entropy relative to the agent here already?}. In the case of collapse, all the correlations must be classical, as the system has already collapsed into one or the other state. By making the Step 2 spin measurement a selective measurement, the Step 4 location measurement becomes redundant.

Since H\&S want to include the non-selective spin-measurement at Step 2, the Step 4 location measurement is indeed a necessary requirement for the single particle case, given that the compression is not being allowed to be conditional on the outcome of the Step 2 measurement. Without the selective measurement, the work-free compression against vacuum could not take place. The limiting case of infinite particles (and in fact von Neumann's original account) does not require this selective measurement since the amount of particles within each chamber of the box becomes equal. 

The problem with including a second measurement on the system is that this second measurement also introduces a second measurement apparatus. It will be shown shortly that H\&S's conclusion is based on an erroneous calculation of the von Neumann entropy when the system is correlated to this second measurement apparatus. Before elaborating on this point however, I would like to discuss another shortcoming of their argument. 
%And, given that it allows us to arbitrarily violate the second law, a severe one moreover. 

%The above considerations show that once we include active manipulations of the system through an external agent that require correlations between the state of the thermodynamic system and the agent's state (such correlations occur during a measurement process), the entropy contribution of the agent herself needs to be taken into account during a cycle. Thermodynamics however is badly equipped to deal with correlations, and hence arguments that aim at saving the second law need to emerge from statistical mechanics, and not thermodynamics (in the words of Earman and Norton's distinction \citep{earman_exorcist_1999}, this is taking the ``sound'' path).

\subsection{Violation of the Second Law}\label{sec:violation}

H\&S  notably claim that the thermodynamic entropy change is zero throughout the whole cycle and in particular, that at the end of the cycle ``[...] the measuring device [can be returned to its initial ready state] unitarily.'' \citep[p.161]{hemmo_von_2006}, and hence without any heat cost. 

To appreciate the consequences of this claim, let us assume that it is indeed possible to unitarily bring the measurement apparatus back to its original position without a compensating heat transfer into the environment, as H\&S claim.
%(against Landauer's principle \citep{landauer_irreversibility_1961,bennett_logical_1973}). 
We may then construct a slightly amended version of their proposed cycle. For this amended version, the only thing we change is Step 5, which instead of being a compression we turn into an isothermal expansion. This means that instead of compressing the empty side against the vacuum, we let the particle push against the partition in a quasi-static, isothermal fashion. Given that the position of the particle is `known' as a result of the location measurement, it is possible to attach a weight to the partition, thereby extracting $kT \ln2$ units of work from the system during the expansion, while the according amount of heat is delivered from the heat reservoir. After the work extraction, the measurement apparatus is brought back to its initial state (which according to H\&S can be done for free). The partition is then re-inserted into the original system (for free), the position of the particle measured again (for free) and the above process is repeated, thereby extracting arbitrarily large amounts of work from this one-particle engine with the sole effect being that heat is extracted from a single reservoir. This constitutes a direct violation of the Kelvin-Planck statement of the second law \citep{planck_treatise_1991}. 

Von Neumann himself, as the authors acknowledge, states that 
\begin{quote}\small\singlespacing
[...] in the sense of phenomenological thermodynamics, each conceivable process constitutes valid evidence, provided that it does not conflict with the two fundamental laws of thermodynamics.'' \citep[p.192]{von_neumann_mathematische_1996}\footnote{[...] im Sinne der ph\"anomenologischen Thermodynamik ist jeder denkbare Prozess beweiskr\"aftig, wenn er die beiden Haupts\"atze nicht verletzt.}, 
\end{quote}

and so at this point, one might already conclude that Hemmo and Shenker's argument fails, as their suggested unitary reset \textit{does} conflict with one of the fundamental laws of thermodynamics. 

It should be noted, however, that there exists some controversy in the literature about whether or not we ought to expect the second law to hold in the single particle case and whether or not thermodynamic considerations in these cases are indeed acceptable\footnote{I am grateful to two anonymous referees for prompting me to elaborate on this point.} \citep{maxwell_diffusion_1878,earman_exorcist_1999,norton_all_2013,hemmo_road_2012}. In fact H\&S elsewhere have expressed skepticism on whether the second law holds in such cases \citep{hemmo_road_2012}. While it seems reasonable to assume that in the article under consideration the second law is required to hold (after all, the article is about comparing the \textit{thermodynamic} entropy with the von Neumann entropy and the authors explicitly state that they ``do not address this issue'' of skepticism here \citep[p.158]{hemmo_von_2006}), it is worth emphasising that the above violation is a \textit{reliable} violation of the second law. This means that even if one is willing to sacrifice the strict second law which states that entropy should never decrease, and instead adapts either a \textit{statistical} or a \textit{probabilistic} version of the second law\footnote{The statistical version, that Maxwell adapted, takes the second law to be a statistical, as opposed to mathematical, truth that holds for macroscopic systems where fluctuations are rare. Naturally, the statistical second law does not hold anymore for microscopic systems where fluctuations become relevant. The alternative is a probabilistic version, which rules out the \textit{reliable} extraction of work from a single heat reservoir. For this probabilistic law, system size seems less of an issue. See also \cite{maroney_information_2009} or \cite{myrvold_statistical_2011} for a distinction between these different versions.}, the problem with the above case is that we are confronted with a violation that allows us to extract work from the heat reservoir 100\% of the time. It allows us to extract work reliably and continuously, and so is a particularly serious version of a Maxwell's demon. 

A common strategy to recover the second law from situations like the one presented above, is to invoke Landauer's principle \citep{landauer_irreversibility_1961,bennett_logical_1973}. It states that there is a heat cost involved with the resetting of the measurement apparatus  to its initial state and is widely accepted in the physics community.\footnote{Notably, some philosophers have challenged its validity \citep{earman_exorcist_1999,norton2011waiting}, while others have made arguments in its favour \citep{ladyman_connection_2007,maroney_generalizing_2009,ladyman2013landauer,wallace2014thermodynamics}} For the case presented above, the resetting of the memory cell would then lead to an increase of heat in the environment, thereby offsetting the previous entropy decrease.\footnote{Note that once we take into account this heat cost, H\&S \textit{original} one-particle cycle ceases to be entropy neutral, as the resetting step will lead to an entropy increase in the environment.} 

And indeed, this is what is required in the present case, for contrary to H\&S's assertion, a unitary reset of the measurement device is \textit{not} possible. To see this, we consider the end of the cycle. The measurement device then is in one of the two mutually exclusive states $\ket{-}_M$ or $\ket{+}_M$. As can be easily seen, there then exist \textit{no} unitary operator that reliably maps the memory cell back to its initial, ready state $\{\ket{-}_M,\ket{+}_M\} \mapsto \ket{ready}_M$. The only way to reset the measurement device unitarily is if one recorded beforehand in which one of the two mutually exclusive states the device is in. To do so, however, one requires a measurement on the measurement device itself, performed by a second measurement device. But then one would want to reset this second measurement device unitarily, too, for which a third device would be needed and so on. Eventually, one would run out of resources and a Landauer type reset, at a cost, becomes unavoidable.

\subsection*{Which Entropy?}

Notwithstanding the above criticism, H\&S's main point is that the von Neumann entropy (as opposed to the thermodynamic entropy) during the Step 4 location measurement \textit{decreases}, and that this gives us reason to reject their conceptual equivalence, still stands, or seems to. 

\begin{quotation}\small 
\noindent As a result of the location measurement, the von Neumann entropy decreases back to its original value. \citep[p.163]{hemmo_von_2006}
\end{quotation} 

In this section I will discuss this claim and in particular I will show that: 

\begin{itemize}
\item[(i)] The physically relevant entity is the joint entropy of system \textit{and} measurement apparatus. It remains the same.
\item[(ii)] It is the system's so-called conditional entropy that decreases during the measurement, not the system's marginal von Neumann entropy.
\end{itemize}

I will now begin with a justification of the two claims. In phenomenological thermodynamics, the joint entropy of two systems is always the sum of their respective entropies. The von Neumann (in the classical case the Gibbs-) entropy on the other hand is generally subadditive and additive only in the absence of correlations between two systems:

\begin{equation}\label{eq:uncorrelated}
H(S,M)\leq H(S)+H(M),
\end{equation}

where $S$ stands for `system' and $M$ stands for `measurement apparatus' or `memory cell'. In the concrete case of the Step 4 location measurement, we can model the measurement apparatus $M$ as a box containing a single molecule and divided by a partition. It can then be in one of two mutually exclusive states, corresponding to the position of the molecule, left ($l$) or right ($r$). We assume it needs to be in a `ready'-state before the measurement, which we chose to be $l$. The entropy is zero, in this case. If we consider the case of collapse, then at the time of the measurement the system will have already collapsed into a spin eigenstate. The correlations between the location of the system and the measurement apparatus are then all essentially classical and the von Neumann entropy before the measurement can be rewritten as:

\begin{equation}
H(S)=-k_B \sum_{s=l,r} p(s) \log p(s),
\end{equation} 

where $p(s)$ is the probability of the system being in the left or right chamber of the box. 

Before the Step 4 location measurement, system and measurement apparatus are not correlated, and their joint entropy is given by $H_3(S,M)=H_3(S)+H_3(M)$, where $3$ is taken to denote `Stage 3', or, in other words, `before the Step 4 measurement'. During the measurement, the memory cell will align itself with the position of the particle and the two systems become correlated. The joint entropy now cannot be expressed anymore as the sum of the individual entropies and instead becomes

\begin{equation} \label{eq:conditional}
H_4(S,M)=H_4(S\vert M) + H_4(M) \leq H_4(S)+H_4(M), 
\end{equation}

with $H(S\vert M)$ being the so-called \textit{conditional entropy}, which quantifies how much $S$ is correlated with $M$ and which is given by 

\begin{align}
H(S\vert M)&=-k_B\sum_{s,m}p(s,m)\ln{p(s\vert m)} \\&=-k_B\sum_{m}p(m)\sum_{s}p(s\vert m)\ln{p(s\vert m)}\\&=k_B\sum_m p(m)H(s).
\end{align}

$p(s)$ and $p(m)$ are the probabilities that system and memory cell are found in macrostate $s=l_s,r_s$ or $m=l_m,r_m$ respectively, $p(m,s)$ is their joint probability and $p(s\vert m)$ the conditional probability, with $H(s)=-k_B \sum_s p(s\vert m)\ln p(s\vert m)$. The conditional entropy is non-negative and is maximal when system and measurement apparatus are uncorrelated, , $0\leq H(S\vert M) \leq H(S)$, in which case Equation (\ref{eq:conditional}) reduces to Equation (\ref{eq:uncorrelated}). It is often considered to be the entropy relative to an agent (in this case the measurement apparatus).

Let us now go back to H\&S's claim that the von Neumann entropy of the system decreases during the location measurement. Does it? The answer is no. What decreases, however, is the \textit{conditional entropy} relative to the measurement apparatus:
 
\begin{equation}
H_3(S\vert M)\geq H_4(S\vert M).
\end{equation}

It reduces to zero, because system and measurement apparatus become perfectly correlated during the measurement. And so when H\&S claim that the system's entropy has decreased, what they \textit{mean} is that the system's conditional entropy has decreased. But the conditional entropy is distinct from the marginal entropy. Re-writing the joint entropy of system and measurement apparatus demonstrates this:

\begin{equation}
H_4(S,M)=H_4(S\vert M)+H_4(M)=H_4(M\vert S)+H_4(S).
\end{equation}

As opposed to phenomenological thermodynamics, which treats systems as black boxes, (classical and quantum) statistical mechanics is able to detect correlations between subsystems, allowing us to mathematically handle the concept of `measurement' in the first place. If we associate entropy with the potential to (reliably) extract work from a system, then the conditional entropy certainly quantifies this ability to a certain extent: a memory cell endowed with an automaton would now be able to (reliably) extract work from the system by allowing it to isothermally expand into the other half of the box, thereby raising a weight. Contrary to an external agent who is not correlated to the particle location. But this is just the ordinary Maxwell's demon scenario\footnote{As mentioned before, Maxwell in 1867 introduced the idea of a ``very observant and neat-fingered being'' (as cited in \citep{maxwell_maxwell_1995}), which was intended to demonstrate that the orthodox second law of thermodynamics could be broken in principle by exploiting fluctuations. In the thought experiment, a box filled with monoatomic gas is divided into two parts by a partition into which a small door is inbuilt. The ``being'', later called Maxwell's demon, controls every atom that approaches the door and either lets the atom pass or not. Since the gas molecules are subject to a velocity distribution, he can decide to only let the fast molecules pass into the one direction and to only let slow molecules pass into the other direction. By doing so the demon creates a temperature gradient, allowing him to violate the second law.} applied to a one-particle setting.

What becomes important for thermodynamic treatments in such a setting, is the joint entropy of system \textit{and} measurement apparatus, as the joint system (ideally) has no correlations with the outside and can thus be treated as a thermodynamic black box. And it turns out that the behaviour of the thermodynamic entropy of the joint system, is exactly mirrored by the behaviour of the von Neumann entropy: the \textit{joint} entropy of system and measurement apparatus does not change during the location measurement, but remains the same: 

\begin{equation} \label{eq:joint}
H_3(S,M)=H_4(S,M).
\end{equation}

And so, to summarise the above: all that changes during the location measurement is the conditional entropy, but neither the joint entropy of the system nor the marginal entropy $H(S)$. Furthermore, the joint entropy, just as the thermodynamic entropy of the joint system, remain the same during the location measurement.

\subsection{No Collapse Scenarios}

Let us now consider the case of no collapse scenarios. In no collapse scenarios, following the measurement in Step 4, the measurement apparatus and the location degree of freedom become entangled. The reduced density matrix after tracing out the decohering environment therefore is given by an inproper mixture due to the neglect of the correlations with the environment. After the Step 4 measurement, the density matrix of the combined system and measurement apparatus is given by

\begin{equation}
\rho^{(4,P+M)}= \frac{1}{2}\left( \ket{+_z}\!\bra{+_z}_P\ket{L}\!\bra{L}_P\ket{L}\!\bra{L}_M+\ket{-_z}\!\bra{-_z}_P\ket{R}\!\bra{R}_P\ket{R}\!\bra{R}_M\right),
\end{equation}

where now $\ket{L}_M$ and $\ket{R}_M$ represent the states of the measurement apparatus. The correlations between the measurement apparatus and the system are of a classical nature and so also in the absence of collapse, the von Neumann entropy of system and apparatus has not changed during the Step 4 measurement.

\section{Conclusion}

This article considers von Neumann's introduction of $-\text{Tr}\rho\ln\rho$ as the quantum mechanical generalisation of thermodynamic entropy. In particular, it shows that an argument raised by \cite{shenker_is_1999} and \cite{hemmo_von_2006} against the equivalence of von Neumann and thermodynamic entropy is problematic because a) their reasoning allows for a violation of the second law of thermodynamics and b) the alleged disparate behaviour in von Neumann and thermodynamic entropy during the Step 4 location measurement is in fact due to a wrong calculation of the von Neumann entropy. It is in fact the system's conditional entropy that decreases during the step, leading to the seemingly disparate behaviour of the two entropies. Finally, the article shows that the relevant quantum entropy, the joint entropy of system and measurement apparatus, remains unchanged during the location measurement and thus exactly mirrors the thermodynamic entropy. 
\newpage
\section*{Appendix: Von Neumann, Entropy and Single Particles}\label{sec:response}

In his original setup, von Neumann introduced a `gas' consisting of individual quantum systems, locked up in boxes and placed in a further, giant box\footnote{Such a setup was first proposed by \cite{einstein_beitraege_1914}}. The `gas' represents a imaginary statistical but finite \textit{ensemble}. The density operator, which he calls the `statistical operator', can only relate to such a \textit{Gesamtheit}.    %$N$ is thereby chosen large enough that statistical irregularities can be considered negligible. 
%For von Neumann, statistical mechanics and thermodynamics are not concerned with the interacting constituents of a single (even if very complicated) mechanical system, but rather with such a \textit{Gesamtheit} of many, independent and non-interacting individual systems \citep[p.191]{von_neumann_mathematische_1996}. 
This means that even in the case of an individual quantum system, von Neumann's argument would remain unchanged: the density operator of this individual quantum system would \textit{still} relate to an ensemble of systems and a system containing a single particle would therefore \textit{still} be modeled as an $N$ particle ensemble. The statistical representations of a) a system containing a single particle, and b) a system containing many particles, are therefore identical. This, however, does not imply that von Neumann denies the meaningful application of thermodynamics to individual particles, quite the contrary: von Neumann explicitly 
%
%While the statistical operator and therefore equally the von Neumann entropy, are not concerned with the interacting constituents of a single (even if very complicated) mechanical system, but instead with an ensemble of many, independent and non-interacting individual systems \citep[p.191]{von_neumann_mathematische_1996}, von Neumann . This however does not mean that von Neumann denies that one can meaningfully apply thermodynamics to individual particles \citep[p.212]{von_neumann_mathematische_1996}. Quite the contrary, he explicitly 
considers the case of a single particle in a box \citep[p.212]{von_neumann_mathematische_1996}. He uses the example to demonstrate that the capacity of an agent to extract work from such a single-particle thermodynamic system depends on the agent's state of knowledge about the position of the particle and therefore adopts what one might call an epistemic interpretation of entropy:
%in order to demonstrate that whether or not an external agent can extract work from the particle, depends on the agent's state of knowledge about the position of the particle. If the observer knows for a fact that the particle is located in the left hand side of the box (partitions notably do not need to be present in the box, at this stage), the entropy of the system is less than if the observer is completely ignorant about the whereabouts of the particle. 
%
%One may be inclined to accuse von Neumann of inconsistency, given that on the one hand entropy appears as a statistical concept applicable only to a \textit{Gesamtheit}, but on the other hand it is possible to assign a thermodynamic entropy to a system containing only one particle in a box. This inconsistency disappears once we take entropy as quantifying an observer's epistemic capacities. Von Neumann explicitly writes about `lack of knowledge' about the position of the single particle. In the quantum case, the system being in a mixed state equally quantifies ignorance, but this time ignorance about which pure state the system is in. 
%
\begin{quote}\small\singlespacing
The temporal variations of the entropy are due to the fact that the observer does not know everything, or rather that he cannot determine (measure) everything, that is in principle measurable. \citep[p.213]{von_neumann_mathematische_1996}\footnote{``Die zeitlichen Variationen der Entropie r\"uhren also daher, dass der Beobachter nicht alles weiss, bzw. dass er nicht alles ermitteln (messen) kann, was prinzipiell messbar ist.''}
\end{quote}

While one may raise several severe objections to this reading of entropy and the density operator, the above nevertheless shows that von Neumann's argument in its original intention not only applies to large (macroscopic) systems, but also for small (microscopic) systems.

\newpage
\bibliography{bibliography2} 
\end{document}